\documentclass{ifacconf}
\DeclareUnicodeCharacter{2009}{\,}
\usepackage{natbib}
\usepackage{graphicx}
\graphicspath{{./graphics/}}
\usepackage{lipsum}
\usepackage{color}
\usepackage{xcolor}
\usepackage{booktabs}
\usepackage{soul}
\usepackage{bm}
\usepackage{amsmath}
\usepackage[short]{optidef}
\usepackage{chemformula}
\usepackage{adjustbox}
\usepackage{derivative}
\usepackage{pgfplots}
\usepackage{rotating}   
\usepackage{subcaption}
\usetikzlibrary{decorations.text}
\usetikzlibrary{shapes.geometric}
\usetikzlibrary{fit}
\usepackage{parskip}

\def\signed #1{{\leavevmode\unskip\nobreak\hfil\penalty50\hskip2em
  \hbox{}\nobreak\hfil(#1)%
  \parfillskip=0pt \finalhyphendemerits=0 \endgraf}}

\newsavebox\mybox

\begin{document}

\begin{frontmatter}

\title{Data Quality Over Quantity: Pitfalls and Guidelines for Process Analytics}


\author[1,2]{Lim C. Siang}
\author[1]{Shams Elnawawi}
\author[3,4]{Lee D. Rippon}
\author[5]{Daniel L. O'Connor}
\author[3]{R. Bhushan Gopaluni}

\address[1]{Burnaby Refinery, BC, Canada}
\address[2]{Georgia Institute of Technology, GA, United States}
\address[3]{University of British Columbia, BC, Canada}
\address[4]{Spartan Controls, BC, Canada}
\address[5]{Control Consulting Inc., MT, United States}

\begin{abstract}
A significant portion of the effort involved in advanced process control, process analytics, and machine learning involves acquiring and preparing data. Literature often emphasizes increasingly complex modelling techniques with incremental performance improvements. However, when industrial case studies are published they often lack important details on data acquisition and preparation. Although data pre-processing is unfairly maligned as trivial and technically uninteresting, in practice it has an out-sized influence on the success of real-world artificial intelligence applications. This work describes best practices for acquiring and preparing operating data to pursue data-driven modelling and control opportunities in industrial processes. We present practical considerations for pre-processing industrial time series data to inform the efficient development of reliable soft sensors that provide valuable process insights.

\end{abstract}


\end{frontmatter}

\section{Introduction}

Data-centric artificial intelligence (DCAI) is an emerging field concerned with the development of high-quality datasets for Artificial Intelligence (AI) and Machine Learning (ML) applications. The traditional approach to AI/ML involves the iterative construction of high-performance models given a fixed dataset. The data-centric paradigm inverts the traditional approach and calls for iterative improvements on the dataset given a fixed model \citep{mazumder2022dataperf}. The success of an AI system is critically dependent on data quality, but ironically, researchers at Google found that data work is often undervalued and disincentivized compared to building novel models and algorithms \citep{sambasivan2021everyone}.

Industrial process plants have amassed billions of data points from decades of high-frequency measurements. However, industrial big data are not necessarily good data due to many potential sources of error, noise, and uncertainty. Process plants in continuous manufacturing industries are designed to operate at steady-state. Extracting useful insights from largely steady-state process data often requires additional information about the physics of the actual process. For that reason, industrial process datasets have been described as data-rich but information-poor \citep{DONG199665}.

What makes a dataset `good'? A good dataset is acquired from reliable sources and prepared in a manner appropriate for both the process opportunity and proposed solution. Data acquisition not only includes working with stakeholders to obtain and contextualize data, but it also involves upstream activities such as using process knowledge to identify the opportunity, investigate the feasibility, and determine the data requirements. Process knowledge is important for many data preparation tasks such as validating the integrity of data, identifying features from first principles, designing visualization strategies, as well as data pre-processing tasks such as cleaning, filtering, normalizing, time-shifting, and segmenting data. The intent of the modelling task will also influence data acquisition and pre-processing decisions.

This article provides a practical guide on data collection and preparation for process analytics and model development, using perspectives synthesized from a cross-functional team of academic and industry practitioners. We arrange the paper into several sections that describe common pitfalls and possible solutions in handling process data. We focus on a broad discussion of best practices in the process industries, rather than algorithm-specific requirements \citep{tsai2018pattern}.


\section{Background}

\begin{figure}[ht]
\centering
\includegraphics[width=\linewidth]{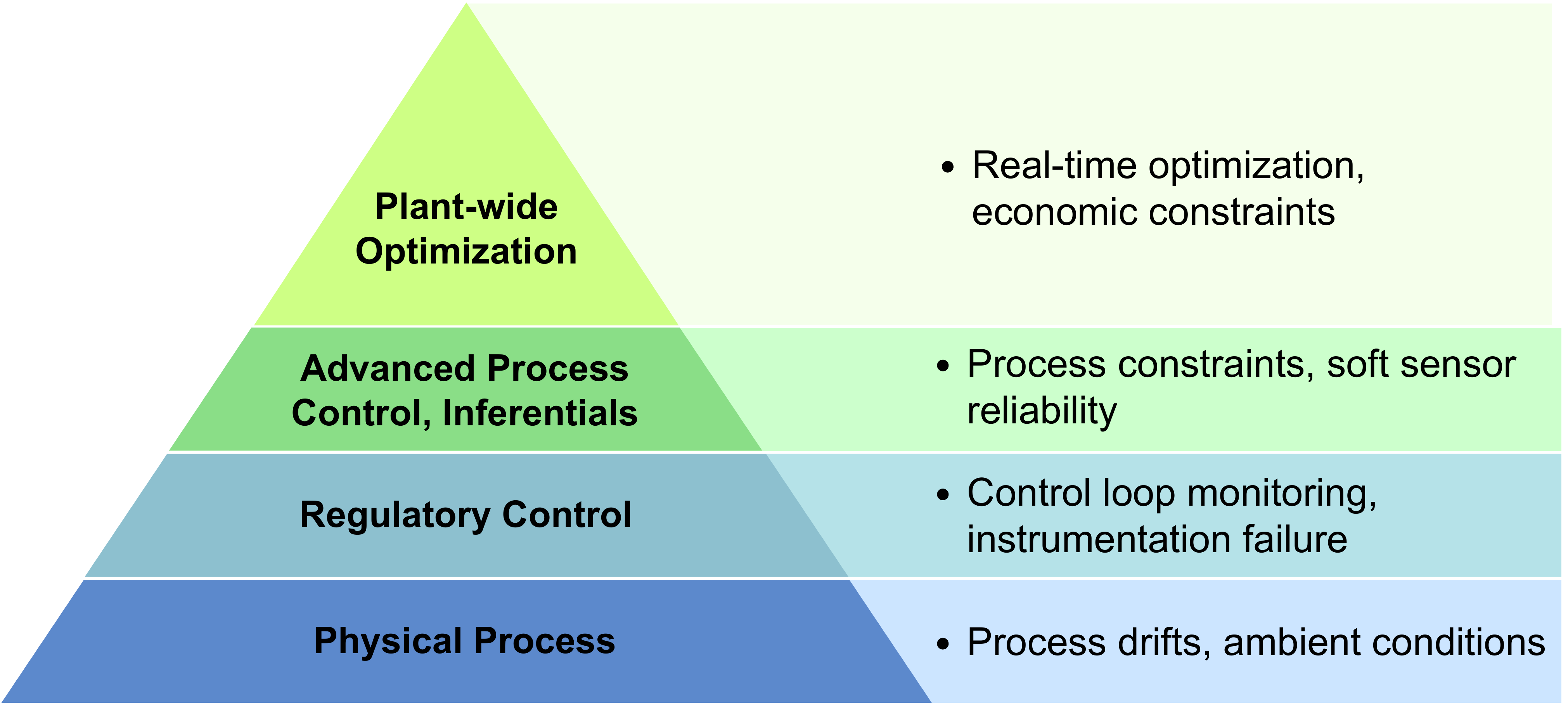}
\caption{Plant process control hierarchy for industrial control systems \citep{hahn2017process}. The text on the right describes factors that strongly affect data reliability within each layer.}
\label{fig:purduePitfalls}
\end{figure}

`Process Analytics' refers to the application of advanced analytics and ML techniques to manufacturing data \citep{sun_opportunities_2020}. The preparation of this data needs to be suitable for the analytics task and desired insights.  From the plant to the end-user, data will undergo a series of steps, each with its own factors that will affect data reliability, as described in Figure \ref{fig:purduePitfalls}.

In this discussion, we use inferential models as an example to demonstrate the practical challenges of data preparation. Inferential models, or soft sensors, are mathematical models used to estimate process quality variables that are critical for control. Economic and physical constraints often limit the ability of implementing on-line sensors and analyzers to provide high frequency measurements of key process variables (PVs). Inferential models are trained with historical data to learn a mapping that generates predictions of key quality parameters from related PVs with reliable on-line measurements \citep{rippon2021representation}.

Inferential modelling frameworks have been proposed by several authors \citep{nian2022simple, kadlec2009data, qin1997neural}. Kadlec (2009) found that many pre-processing steps such as handling missing data, outliers, feature selection, are all done manually. Practitioners have reported that dealing with these data pre-processing steps can take longer than building the model itself \citep{qin1997neural, kano2010state}. Due to the safety-critical nature of industrial operations, control practitioners must have a strong understanding of their data pre-processing needs based on domain knowledge. Engineers typically clean and pre-process data in a transparent, deterministic and interpretable manner. The painstaking preparation of useful data is warranted by the significant influence data quality has on the success and sustainability of industrial data science applications. Recent research has also attempted to automate data pre-processing steps using reinforcement learning \citep{berti2019learn2clean}, but such techniques are not widespread in industrial practice because an explainable model is important for gaining stakeholder acceptance to trust and use it \citep{barton2022model}.

Many academic papers describe the soft sensor modelling steps in detail, discussing various hyperparameter optimizations, evaluating the proposed model's performance compared to competing algorithms, and discussing the model's advantages. However, discussions on how the input data to the model is obtained and managed is often absent, aside from cursory and often vague statements of ``data cleaning'' or ``data pre-processing'' steps like removing 3-sigma outliers or performing normalization on the input data prior to modelling. 


How was the input data retrieved? How were the relevant variables selected and how was the data cleaned? How were the timestamps for multi-rate data aligned? These challenges are resolved behind the scenes between data scientists, plant engineers, Subject Matter Experts (SMEs), operators, and instrumentation technicians. Although these details are essential for experimental reproducibility, they are often not presented in a thorough and comprehensive manner in the final publication.

In a series of experiments, \cite{zliobaite2012adaptive} show how different data pre-processing treatments affect model performance under different types of concept drifts, quantitatively demonstrating the importance of data pre-processing on the final model. Despite the importance of data preparation, there is relatively sparse guidance in the literature on how to handle process data for analytics and ML in a practical manner. In what follows, we provide insights on this topic that are framed as possible pitfalls that may plague the practitioner.

\section{Guidelines for Process Data Analytics}
\label{s3}

Distilling operational insights and production value from industrial process data is essential for sustainable capitalization of natural resources. Here we present five common pitfalls for researchers and industrial practitioners. Each pitfall is accompanied by a discussion on how it can be identified and avoided.

\subsection{Pitfall 1: Failure to correctly retrieve and contextualize process data}

Unlike the contrived datasets widely available for data science or ML benchmark problems that are already applicable for modelling, the data acquisition workflow for industrial processes is a non-trivial step. Due to the large data volume that is retrievable from industrial data historians, it is helpful to frame the data acquisition question not by asking ``\textit{what data do I have?}" like in static, contrived datasets, but rather, by asking ``\textit{what data do I \textbf{want}}?" Therefore, acquiring good industrial process data is not merely a simple transfer of data, but a more complex, iterative exercise with collaborative efforts between data practitioners and plant personnel. Some variables in the data may be irrelevant, or even detrimental, to the modelling task, so appropriate curation of variables based on domain knowledge is necessary for good model development. For example, if a lab sample is only updated weekly, retrieving hourly data for that sample will be misleading and does not reflect the true data resolution, and if a control loop is always running in manual mode, adding a constant setpoint to the dataset will not be meaningful.

\subsubsection{Data access}

Given the iterative nature of industrial data acquisition, having data that is readily accessible to partners is an important step to reduce the cost of data exploration. In joint academic-industry projects, it is more productive to provide direct historian data access to the academic partners involved, so that they can query and acquire the datasets that they need, while the plant engineers can provide support with process advice and data context. Understandably, depending on company policies, such open arrangements are not always possible.

\subsubsection{Data exploration}

Exploratory data analysis (EDA) must be performed to inspect and understand the available data. Appropriate representation and visualization of time series data is important during EDA. The role of data visualization for effective collaborations is severely underappreciated, particularly in the context of using visualization tools to empower operators and SMEs and leverage their domain expertise in data-driven decision-making \citep{rippon2021visualization, ELNAWAWI2022105056}.

The practitioner should always review a sample of the data before attempting to do any large-scale queries. For example, when retrieving data for a plant feed rate, it is logical to remove periods when the unit was shut down. Simple spreadsheet filters or textual queries in SQL may be unsuitable for this task. For example, using a simplistic SQL \texttt{WHERE} clause to remove values below a threshold in the entire dataset, without closer inspection of the underlying time series, will inadvertently remove anomalies outside of the shutdown window that could be indicative of process upsets or instrumentation issues that may actually be useful data.

Most industrial plants have self-service data visualization tools such as Seeq, Spotfire, PI Vision, or CCI Cake \citep{siang2022} that can help make EDA or other data visualization and cleaning tasks easier to perform. Custom-built tools may be warranted for more complex analysis, but the practitioner should consult plant personnel before attempting to develop a costly, time-consuming solution from scratch if existing tools are readily available and satisfy the practitioner's needs.

\subsubsection{Data resolution}

How should we structure our historian queries? Should we retrieve raw, un-gridded data or should we retrieve gridded and interpolated data at intervals of seconds, minutes, hours, or days? If interpolation is desired, do we assume a zero-order hold between data points, or a first-order linear interpolation? These are questions that may be tag or unit-specific, and should be discussed with plant personnel after the initial data exploration work is complete. Treating all tags equally with the same retrieval settings would likely lead to a poorly-constructed dataset due to a mismatch between the data retrieval frequency and the actual data collection frequency. If the process response is slow, collecting high-frequency data will result in an unnecessarily large dataset that may not provide any additional modelling benefits.

Data compression may be a major pitfall in analyzing data from industrial process data historians. \cite{thornhill2004impact} is an excellent quantitative study on industrial data compression that describes the issue in detail and provides guidance for dealing with compression problems. Due to the reduced cost of computing storage over the past two decades, many facilities have deactivated data compression settings in their historians. However, care must be taken when extracting large, multi-year datasets as the older time slices may suffer from compression issues even though the new time slices are unaffected. 

\subsubsection{Data contextualization}

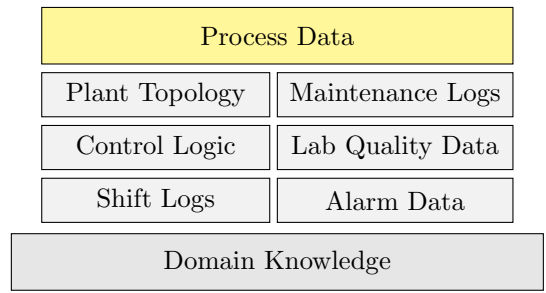
\begin{figure}[]
\centering

\begin{tikzpicture}[every fit/.style={inner sep=0pt, outer sep=0pt, draw}]

\begin{scope}[y=0.75cm]
\node [fill=yellow!50, fit={(0.4,0) (6.6,1)}, label=center:{Process Data}] {};
\end{scope}

\begin{scope}[yshift=-0.7cm,y=0.65cm]
\node [fill=black!5, fit={(0.4,0) (3.4,0.9)}, label=center:{Plant Topology}] {};
\node [fill=black!5, fit={(3.5,0) (6.6,0.9)}, label=center:{Maintenance Logs}] {};
\end{scope}

\begin{scope}[yshift=-1.4cm,y=0.65cm]
\node [fill=black!5,fit={(0.4,0) (3.4,0.9)}, label=center:{Control Logic}] {};
\node [fill=black!5,fit={(3.5,0) (6.6,0.9)}, label=center:{Lab Quality Data}] {};
\end{scope}

\begin{scope}[yshift=-2.1cm,y=0.65cm]
\node [fill=black!5,fit={(0.4,0) (3.4,0.9)}, label=center:{Shift Logs}] {};
\node [fill=black!5,fit={(3.5,0) (6.6,0.9)}, label=center:{Alarm Data}] {};
\end{scope}

\begin{scope}[yshift=-3cm,y=0.75cm]
\node [fill=black!10,fit={(0,0) (7,1)}, label=center:{Domain Knowledge}] {};
\end{scope}

\end{tikzpicture}

\caption{Process data must be contextualized with other datasets to provide meaning.}
\label{fig:processdata}
\end{figure}

As shown in Figure \ref{fig:processdata}, process data are one of many sources of plant information. Plant topology data from P\&IDs provide connectivity and causality information, informing the practitioner on how process tags are related. Domain knowledge from operating standards and discussions with plant personnel can provide important first-principles information about the underlying physics and chemistry of the process. Maintenance records can provide instrumentation reliability information to determine the trustworthiness of the measurements. Control logic diagrams allow the practitioner to interrogate the underlying data generation process for a historized tag. As facilities modernize and undergo digital transformation, less effort is required by the practitioner to collect, parse and understand these disparate information sources.

Data practitioners should be aware that tag names and labels may be misleading. For example, one might assume that the tag historizes a raw measurement signal, but in reality, the tag could be a calculated value that applies transformations like filtering, clipping or splicing to the signal. This can be reviewed by inspecting the distributed control system (DCS) logic diagrams or historian calculations. Lab data can provide a ground truth for soft sensors, but it is often collected at a much slower rate. Finally, shift turnover logs and alarm data highlight abnormal plant conditions that should be managed accordingly.

\subsubsection{\textbf{Summary 1}} EDA is an essential step for understanding the available data. Practitioners need to be aware of the contexts from which they pull and use process data, and they must ensure that data retrieval is suitable for the task at hand. Practitioners should consult plant personnel to determine if existing visualization tools can be used to avoid unnecessary effort. Engaging in discussions with the facility's historian administrator or process control engineers can help the practitioner understand potential DCS logic or compression issues. Lastly, different sources of plant data like P\&IDs, control logic diagrams, operating standards, and maintenance records should be examined to obtain a holistic understanding of the data being used for further analysis.

\subsection{Pitfall 2: Ignoring domain knowledge and assuming that data volume and variety can make up for data quality}
Industrial big data are not always good data. A common misconception is that data volume and variety can make up for data quality. Continuous chemical processes operate at steady-state by design, in contrast with more dynamic systems like robotics or autonomous vehicles. Adding more features and more data volume will not necessarily improve model performance, as demonstrated in prior work where the benefits of new features are outweighed by the curse of dimensionality \citep{rippon2021representation}. It is also well-recognized in the literature that a purely data-driven, black-box soft sensor model that disregards the physics of the plant will be unreliable when the process operates under conditions that were not captured during data collection \citep{kano2010state}.

\subsubsection{Data cleaning}

Data points outside their `normal' operating range should be discarded. Statistical techniques like a 3-sigma outlier removal procedure may be simple to apply, but these techniques may not be effective if the process has different modes of operation. Xu et al. (2015) distinguish between model-based and data-driven outlier detection techniques, and further describe outliers as either univariate or multivariate depending on how many components of the data are part of the outlier. As an example, the 3-sigma rule postulates that outliers can be detected if the following condition is true:

\begin{equation}
    \left| x_t - \mu \right| > 3\sigma
\end{equation}

where $x_t$ is the data point being tested, $\mu$ is the mean, and $\sigma$ is the standard deviation of the dataset. However, the mean and standard deviation will change depending on different modes of operation. It may be more effective to use model-based outlier detection or techniques with a moving time window to account for different operating modes \citep{xu_data_2015}. 

\begin{figure*}[t!]
    \centering
    \begin{subfigure}[t]{0.5\textwidth}
        \centering
        \includegraphics[height=2.0in]{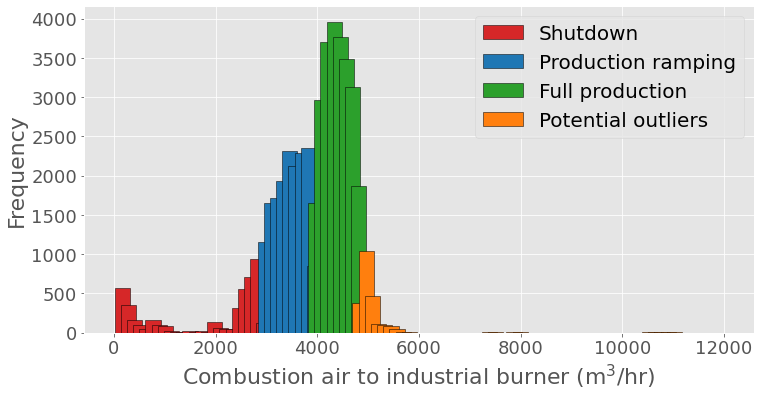}
        \caption{Histogram of process data with operating categories.}
    \end{subfigure}%
    ~ 
    \begin{subfigure}[t]{0.5\textwidth}
        \centering
        \includegraphics[height=2.0in]{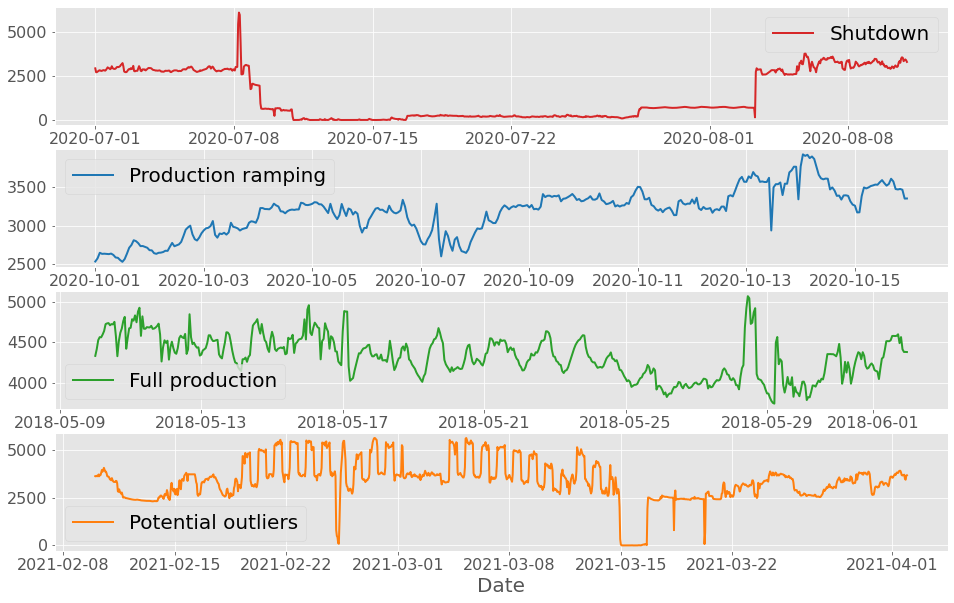}
        \caption{Examples of categories of operating data.}
    \end{subfigure}
    \caption{Common characteristics of process data are illustrated as (a) histogram partitions and (b) time series examples.}
    \label{fig:hist}
\end{figure*}

Figure \ref{fig:hist} presents industrial data to demonstrate an archetypal example of a process value partitioned into four common operating modes. Figure \ref{fig:hist}a (left) shows a histogram of combustion air flow rates and Figure \ref{fig:hist}b (right) shows sampled periods of time series data corresponding to each mode of operation. Many PVs share a highly similar distribution which can help simplify data cleaning, but it is important to note that the distribution of some PVs can be significantly different. Normal operating ranges for various process data can be obtained by consulting plant personnel with \textit{a priori} knowledge \citep{shang2014data}. 

Common heuristics used by practitioners are listed below. Data that do not conform to these heuristics should be reviewed for potential issues, depending on the application.

\begin{itemize}
    \item Check that the tags of interest are actually moving beyond typical background noise levels. A static input that has no significant movement or excitation in the data provides no predictive power. A tag that does not move at all could be indicative of instrumentation or historian issues.
    
    \item Check for multicollinearity. If two inputs are collinear, such as redundant temperature sensors, depending on the modelling algorithm and intent, it is unlikely that both are needed. Adding more inputs to the model will create more points of failure that can degrade model reliability.
    
    \item Check that the regulatory control system is functioning well and the control loops are well-tuned. If the control loops are oscillating, the data can be detrimental for model performance.
    
    \item Check for mass and energy balance closure and mis-calibrated sensors.
\end{itemize}

During data cleaning, the practitioner must identify whether the input measurements are sensible. Industrial instrumentation can be faulty due to maintenance or calibration issues. Sensors can be calibrated for past operating conditions that are no longer valid and they may have varying quality under different process conditions. Plant personnel and asset reliability data can help identify instrumentation issues. For a systematic and data-driven approach, data reconciliation and gross error detection should be considered. In particular, an accurate mass balance may be critical if the application involves regulatory compliance, such as carbon accounting. For an in-depth treatment of these topics, the classical text by \cite{narasimhan1999data} is an excellent resource.

\subsubsection{Feature selection}

P\&IDs should be thoroughly reviewed to identify key manipulated variables (MVs) and controlled variables (CVs) in a process before any data-driven modelling is attempted. Tags that are indicative of different operating conditions should be identified. For instance, in pulp mills, valve positions can indicate the use of different process fluids (e.g., filtrate, mill water, condensate, or white water). Refining processes are typically locally-linear, and multiple linear models might be needed to cover different process states. Some product specifications are seasonal, e.g., higher Reid Vapor Pressure (RVP) gasoline blends are produced in the winter. Column tray temperatures are indicative of product composition, and a pressure-compensated temperature (PCT) is commonly used for composition prediction. Observing manual operator interventions in response to off-spec products can also help identify key MVs and CVs in the process that may not be present in the control scheme.

Isolating relevant features can also include making numerical changes like normalization, biasing, or conversion of raw data. An iterative approach may be required to identify the optimal set of features. Computational techniques such as the Shapley value \citep{cao2022soft} and causal graphs \citep{cao2022causal} could also help find relevant features and can be a powerful approach when combined with domain knowledge.

\subsubsection{Steady-state and transient data}

Industrial plants in continuous manufacturing are designed to operate at steady-state to ensure reliable and consistent production. The process is in a `transient state' when it moves between different operating points. Both steady-state and dynamic analyses are important in process control applications, so being able to distinguish them in process data will help practitioners develop better models. For dynamic models, practitioners must ensure that training data includes transient data to capture process dynamics.

Numerous methods for identifying steady and transient states both online and offline have been developed over the years \citep{wu2015online, kelly2013steady}. As a practical example of these techniques, Rhinehart developed a statistical method based on the ratio of two variances that is computationally simple and suitable for online implementation called the R-statistic \citep{cao1995efficient}. The R-statistic is near unity when the process is at steady-state, and can be compared to a `critical value' to determine if the null hypothesis of being at steady state can be rejected \citep{rhinehart2013automated}.

Calculating the R-statistic on key process variables such as feed rate can guide practitioners to systematically select appropriate steady-state and transient-state sections of data for model development. During system identification, extended periods of steady-state data absent of any process excitations should be removed, as those regions do not contribute any useful dynamic modelling information.






\subsubsection{\textbf{Summary 2}}
Practitioners must understand \textit{which} subset of
data are necessary for the analysis tasks involved, both in terms of features and time windows. Practitioners should only use data within the appropriate ranges, and these will depend on which operational modes are important for the type of analysis performed. This includes many factors, such as posturing units for making different products or identifying steady- and transient states during operation. Data should be checked for adequate variability in the selected ranges, and multivariate factors like multicollinearity and mass/energy balance closures must be properly understood. Some variables may consist of combinations of others like PCTs and RVPs, and the underlying inputs should be reviewed carefully.

\subsection{Pitfall 3: Failure to differentiate between open-loop and closed-loop conditions}

One distinguishing characteristic of industrial process data is the presence of closed-loop operating conditions under feedback control, which is rarely found in other datasets. It is well-recognized that empirical models fit with closed-loop data instead of open-loop data will show a sign change in the coefficients \citep{kresta1994development}.

Although closed-loop identification methods can build open-loop correlations from closed-loop data, practitioners need to be aware of the effect that feedback has on a given dataset. Closed-loop data introduce correlations that are not present in open-loop data. MacGregor provided a simple example; consider a simple distillation column with automatic control of overhead purity using reflux rate. Distillation fundamentals tell us that increasing reflux should increase overhead purity, and vice versa. However, under closed-loop control, the data showed a negative correlation, because both operator intervention and the controller are responding to low overhead purity by increasing the reflux, and vice versa \citep{macgregor1991some}. If the practitioner were to ignore domain knowledge and chemical engineering fundamentals, they would erroneously `discover' that higher reflux leads to poorer product separation based on trends in the closed-loop data.

For more practical considerations, how can we detect the presence of open-loop and closed-loop conditions in the data? For a single PID loop, one can query the data historian for the loop mode tag, typically prefixed with \texttt{.MODE} or \texttt{.AM} (Auto/Manual). For multivariable control and APC systems, one could query the data historian for the APC systems' \texttt{ON} status. Furthermore, non-critical APC variables may have their service status turned off for maintenance or other operational reasons, even though the overall APC system is active. For these reasons, it is important to study and understand the underlying control logic that governs the tags, especially if more advanced control techniques like multivariable control, cascade control or gap control are used.

The relationship between input variables (i.e., MVs) and output variables (i.e., CVs) is governed by process dynamics. Changes in the MVs require time to propagate through the system before the CVs return to steady-state. This time is known as the `time to steady state' (TTSS) in industrial model predictive control (MPC) terminology. If steady-state modelling is performed without using traditional system identification techniques, \cite{ferreira2022development} stressed the importance of applying a time shift to align the input and output signals and noted a lack of literature discussing this issue. This time shift is akin to applying the appropriate time delay to the signals. In practice, this time delay will likely be a time-varying time delay, as feed rate changes will affect the residence time of material in the process. Failure to apply the appropriate time shift can lead to poor steady-state modelling.

Sensor data can be manually aligned by estimating the residence times using domain knowledge. If the facility is utilizing an APC system with reasonably accurate models, they should provide a good estimate for the TTSS. If the relationships under study are not configured in the APC models, then system identification techniques can be applied to historical process data to probe these relationships. If historical data is largely steady-state and does not meet the practitioner's needs, a step test could be performed in collaboration with plant personnel to discuss what moves to make and which control loops to use.

\subsubsection{\textbf{Summary 3}}
Practitioners should be cognizant of open- and closed-loop conditions when acquiring data. Models can be built using data from both modes of operation, but the chosen modelling techniques must take this into account. As illustrated by MacGregor's distillation control example, data under closed-loop conditions may show counter-intuitive effects and sign changes. If the data acquisition tasks require step testing, engineers need to create a detailed test plan to ensure proper collection of data that minimizes hindrances to normal operation.

\subsection{Pitfall 4: Failure to align multi-rate data correctly}

Industrial processes are multi-rate systems because a large number of online measurements are sampled at a fast rate (e.g., seconds), but their corresponding quality variables, measured in a laboratory, are sampled at a much lower frequency (e.g., hours or days). We illustrate this multi-rate data alignment issue with a classical example of reconciling process and lab readings in estimating the bias for soft sensor models. 

Soft sensor performance can be measured by calculating residuals, whereby model predictions are compared to the `true' values using lab quality variables measured offline. These lab quality variables are obtained with significant time delay, e.g., hours or even days after sample collection. Due to this time delay, intuitively, the lab results must be compared with the inferential prediction at the sampling time as opposed to the time when the lab results are returned \citep{lu2018semi, wang2019monitoring}.

\begin{figure}[h]
\begin{center}
\includegraphics[width=8.4cm]{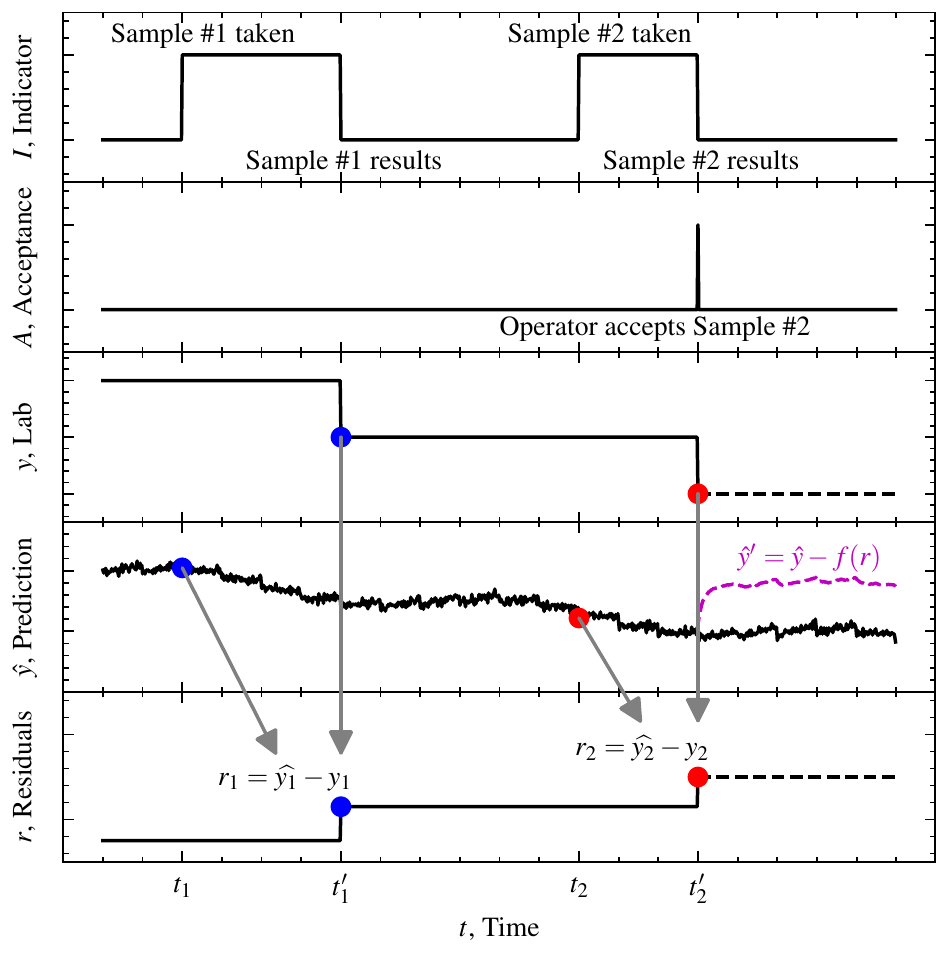}    
\caption{Lab sample collection and time shift methodology.} 
\label{fig:labshift}
\end{center}
\end{figure}

We wish to compute the residual, $r_i = \hat{y_i} - {y_j}$ where $y_j$ is the lab quality results available at time $t_j$ from a sample collected earlier at time $t_i$, and $\hat{y_i}$ is the model prediction at time index $t_i$, as shown in Figure \ref{fig:labshift}. There is a need to track the sampling time and align the time stamps of the lab results with the inferential predictions for correct computation of the residuals. In industrial plants, a common configuration in the DCS logic is to provide a lab sample switch for operators to control a binary indicator function, $I$. When a sample is taken, the operator will toggle the sample switch which changes $I$ from 0 to 1. When the lab results are available, $I$ will reset from 1 to 0. Therefore, $I$ provides a mechanism to systematically track $t_i$ and $t_j$ that are required for the residual calculations.

In reality, the lab results received may not be correct due to human errors or equipment malfunctions. The operator must use domain knowledge and experience to judge the trustworthiness of these lab results. In industrial plants, an additional layer of data validation is typically provided in the form of an operator toggle switch. This switch triggers an acceptance pulse, $A$, if the operator accepts the lab results. This concept is illustrated in Figure \ref{fig:labshift}, where the first lab result at $t'_1$ is rejected, and the second lab result at $t'_2$ is accepted because a pulse is present.

An operator-accepted residual is used as a bias term, $f(r)$ to correct the raw inferential predictions, such that the corrected prediction is $\hat{y}' = \hat{y} - f(r)$. The bias is determined by the most recent lab sample accepted by the operator, which is $r_2$ in this example, as shown by the purple line in Figure \ref{fig:labshift}. The function $f$ applied to the residual indicates that the raw residual value might not be used directly for control. In industrial practice, a lag filter is usually applied on the bias to avoid making large changes to the inferential and bumping the controller.

One common issue that is often overlooked is the uncertainty in sample collection time, $\Delta t$. The actual sample collection time, denoted by $t_i' = t_i + \Delta t$, can be affected by the physical location of the sampling point. The uncertainty may be particularly large if the sampling point is located far from the process unit, if the piping configuration involves a long circuit with deadlegs, or if the sample flow rate is slow. A large $\Delta t$ could also stem from incorrect usage of the manual sample collection indicator switch, $I$, such as when the switch is flipped too early or too late relative to taking the sample. If the process is truly at steady-state, $\Delta t$ may not have a significant effect, but nonetheless, having a thorough understanding of how the sample timestamps are generated will help the practitioner construct a high-quality dataset.

In facilities that use hardware and equipment from multiple vendors, each with their own independent data collection systems, care must be taken to synchronize their internal clocks and data timestamps. Failure to keep these clocks synchronized will result in temporal uncertainties in the data across different equipment and areas of the plant. These uncertainties will lead to difficulties in performing a meaningful plant-wide analysis, and worse, may even invalidate results and conclusions.

\subsubsection{\textbf{Summary 4}}

Practitioners must consider how multi-rate signals are cleaned and aligned in their datasets. Domain knowledge can be incorporated in data cleaning by discarding lab samples that are not accepted by the operator. The prediction-correction scheme must be understood in the control logic because the bias term will likely have filters or other transformations applied to it. The physical location of the sampling point and its configuration should be inspected and properly accounted for to accurately estimate sampling times. If equipment and hardware from multiple vendors are used, care must be taken to properly synchronize their timestamps.

\subsection{Pitfall 5: Chasing irrelevant model metrics at the expense of business and environmental outcomes}

Soft sensors often fail to provide sustained value in industrial production environments after initial commissioning. The following list contains a subset of the many possible reasons why a soft sensor model can fail:
\begin{enumerate}
    \item Process drifts - conditions can shift over time due to feed quality, equipment changes, fouling, etc.
    \item Instrumentation issues - instruments such as pressure taps may get plugged up due to fouling or cold conditions and provide incorrect readings.
    \item Sampling station issues - if the sample line is too long and the sample flow is too slow, the line must be thoroughly flushed to collect a representative sample.
    \item Lab equipment issues - results from lab devices may drift may slowly drift away from their true value and a sharp drop in error will be apparent after calibration.
    \item Regulatory control issues - samples collected when the process is not at steady-state may reflect transient process conditions that may not be suitable for the desired modelling objectives.
\end{enumerate}

A 2010 survey revealed that over 90\% of industrial soft sensors in Japan use linear modelling techniques like multiple linear regression and partial least squares \citep{kano2010state, kim2013long}. Depending on the sophistication of the DCS at the facility, implementing a complex model like a deep neural network may not be feasible without costly infrastructure upgrades. Simpler models may be preferable if they help drive business value \citep{nian2022simple}. The survey also noted that the most pressing problem with industrial inferential sensors is not model accuracy, but rather, model maintenance. How do we design durable soft sensors that maintain high accuracy for long periods of time given the dynamic nature of the underlying process? These issues must be given due consideration during model development.

For real-world industrial problems, model metrics such as mean absolute error (MAE), root mean squared error (RMSE), or coefficient of determination ($R^2$) are part of a broader set of consideration such as development time, durability and interpretability. The practitioner should be mindful of the 80/20 rule or Pareto Principle during model development and be aware that striving endlessly towards performance metrics may come with rapidly diminishing returns. For example, trying to improve a model's accuracy from 80\% to 90\% will likely be significantly easier than squeezing out an extra 0.1\% of performance to go from 99\% to 99.9\%. Furthermore, the presence of sensor noise, measurement errors, uncertainties in lab results, and other practical process-specific considerations may significantly outweigh minor performance improvements in model accuracy.

Unless there are significant business benefits, spending excessive time to strive for perfection in model metrics may not be the most productive use of the practitioner's time, nor be in alignment with the priorities of the business stakeholders. From an operational perspective, the practitioner should understand how the models they are building will be used to drive plant improvements. It is important to consider the business impact and intended usage of the models, and not treat model development as merely an academic exercise to maximize certain metrics.

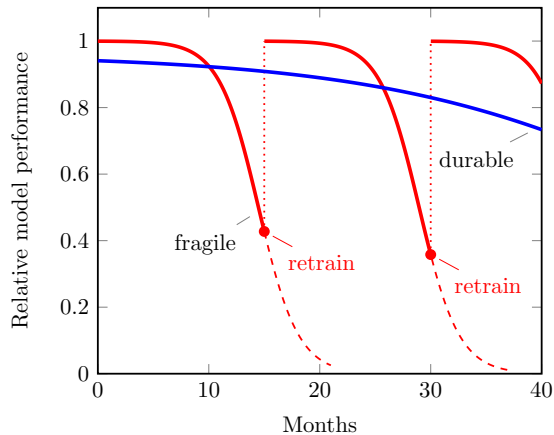
\begin{figure}[]
\centering
\resizebox{0.85\columnwidth}{!}{
\begin{tikzpicture}
[
  declare function={func(\x,\a,\k,\b) = (\x < \k)*(-0.5*tanh(\a*\x-\k/\b)) +
                                        and(\x > \k, \x < 2*\k)*(-0.5*tanh(\a*\x-2*\k/\b)) + 
                                        (\x > 2*\k)*(-0.5*tanh(\a*\x-3*\k/\b)) +
                                        0.5;
                    resid(\x,\a,\k,\b)= and(\x > \k, \x <2*\k)*(-0.5*tanh(\a*\x-\k/\b)) + 
                                        (\x > 2*\k)*(-0.5*tanh(\a*\x-2*\k/\b)) +
                                        0.5;        
                    durable(\x,\a,\k,\b) = (-0.48*tanh(\a*\x-\k/\b)) + 0.48;
                   }
]
\begin{axis}[samples = 501, 
             ymin=0, ymax=1.1,
             xmin=0, xmax=40,
             domain=0:40, thick,
             x label style={at={(axis description cs:0.5,0)},anchor=north},
             y label style={at={(axis description cs:0.05,0.5)},rotate=0,anchor=south},
             xlabel = Months,
             ylabel = Relative model performance]
             
            \addplot [domain=0:15,mark=none,draw=red,ultra thick] {func(x,0.28,15,3.7)} node[pos=0.97,pin={[pin distance=0.10cm]below left:{$\text{fragile}$}}] {};
            \draw[dotted,red] (axis cs:15,{func(14.9999,0.28,15,3.7)}) -- (axis cs:15,1);
            \addplot [domain=15.01:30,mark=none,draw=red,ultra thick] {func(x,0.28,15,3.7)};
            \draw[dotted,red] (axis cs:30,{func(29.9999,0.28,15,3.7)}) -- (axis cs:30,1);
            \addplot [domain=30.01:40,mark=none,draw=red,ultra thick] {func(x,0.28,15,3.7)};
            
            \addplot [only marks,color=red,samples at={14.9999}] {func(x,0.28,15,3.7)} node[pin={[pin edge={red},pin distance=0.15cm]below right:{retrain}}] {};
            \addplot [only marks,color=red,samples at={29.9999}] {func(x,0.28,15,3.7)} node[pin={[pin edge={red},pin distance=0.15cm]below right:{retrain}}] {};
            
            \addplot [domain=15.01:21,mark=none,draw=red,dashed] {resid(x,0.28,15,3.7)};
            \addplot [domain=30.1:37,mark=none,draw=red,dashed] {resid(x,0.28,15,3.7)};
            
            \addplot [no markers,
                      draw=blue,
                      ultra thick] {durable(x,0.034,15,7.7)} node[pos=0.99,pin={[pin distance=0.10cm]below left:{$\text{durable}$}}] {};
\end{axis}
\end{tikzpicture}
}
\caption{Durable inferential sensors reduce the burden of model maintenance.}
\label{fig:durable}
\end{figure}

For model updates, it may be useful to consider the new model's performance not in absolute terms based on its RMSE or $R^2$ value, but rather, in relative terms compared to the baseline performance. If the existing model is poor, even a new model with modest RMSE or $R^2$ values could be valuable. Simple, interpretable algorithms may also work well and should be considered first. Achieving some quick wins will demonstrate business value and a clear return on investment, leaving the practitioner free to explore more advanced solutions while providing operating value and keeping stakeholders satisfied.

Key performance trade-offs must be considered during model development, such as production objectives, model maintenance, and operator training. \textit{Model durability} is the length of time a model is in production before performance degradation necessitates tuning or retraining. As illustrated in Figure \ref{fig:durable}, a high-performing model may not necessarily be a durable model. A durable model, even with a slightly lower performance relative to a fragile model, requires less frequent maintenance and would typically be preferred by plant personnel. The production losses incurred by an inferential model being unreliable or unavailable may be several orders of magnitude greater than the opportunity cost of a model operating at a slightly lower accuracy until the next update. Model metrics like the prediction error are only a small part of delivering value in practice, but unfortunately, the soft sensor literature focuses on predictive performance \citep{barton2022model}, and not enough attention is given to practical issues like model durability and interpretability.


\subsubsection{\textbf{Summary 5}} 

Practitioners must acknowledge that in industry, regression metrics (e.g., RMSE) are considered in a broader context including robustness and maintenance. Despite the increasing sophistication of ML tools, the implementation of such algorithms is limited by the hardware capabilities in the plant, and complex models may not be the most appropriate or even feasible solution. Unlike other domains, models developed in a safety-critical industrial plant must be interpretable for plant personnel to understand, trust and maintain them.

\section{Conclusions}

In process analytics, exploratory data analysis is essential. Blindly feeding data into algorithms without understanding the reliability and quality of the data can result in poor and fragile models with low acceptance and trust from plant personnel. While data pre-processing and collection are understood heuristically within organizations, there are many fundamental open questions to address on this topic. How can we improve reconciliation of data from different sources? Can we ensure consistent model performance under process changes and concept drifts? Can we capture and generalize lessons in data collection and preparation? Can we automate data acquisition and pre-processing, or develop a systematic methodology for soft sensor development? Such research directions may provide significant unrealized benefits to the broader chemical engineering and process control community. This work provides guidance for practitioners to navigate common pitfalls encountered during data acquisition and preparation, which in practice, have been resolved by \textit{ad hoc} discussions between the data practitioner and experienced plant personnel, but often undocumented in the literature.

\bibliography{citations}

\end{document}